# Accurate calculation of field expansion coefficients in FEM magnetostatic simulations

Vasily Marusov

*Abstract*—FEM simulations are a standard step in the design of accelerator magnets. It is custom for accelerator applications to characterize the field quality in terms of field expansion coefficients. With a commonly accepted approach, expansion coefficients are calculated by means of a Fourier transform of the local FEM solution at points on an arc. The accuracy of the coefficients calculated this way depends strongly on the FEM mesh configuration and simple refinement of the mesh does not always improve accuracy. The accuracy of the expansion coefficients calculation can be improved by using the data on the magnetization of elements in the magnet yoke, obtained in the solution, instead of using directly the local solution. Since currents and the yoke magnetization are the only sources of the field, with these data the field expansion coefficients can be calculated at any remote point. We derive closed forms for calculating expansion coefficients and implemented these results in the ANSYS® add-on. Results for a case study are presented, which demonstrate that expansion coefficients can be calculated with good accuracy even for a rather coarse mesh.

## I. Introduction

FOR most practical purposes, the magnetic field analysis for accelerator applications can be reduced to a two-dimensional problem and the field can be represented by the power series expansion

$$\boldsymbol{B(z)} = B_y + i B_x = \sum_{n=1}^{\infty} \underbrace{(B_n + i A_n)}_{\equiv C_n} \left(\frac{\boldsymbol{z}}{R_{ref}}\right)^{n-1} \quad (1)$$

Here $B_x$ and $B_y$ are components of the magnetic flux density, $\boldsymbol{z} = x + i y$ and $R_{ref}$ is a reference radius. The complex expansion coefficients $\boldsymbol{C_n} = B_n + i A_n$ in commonly accepted jargon are called "multipoles" or "harmonics".

Representation of the field in the form of a power series is possible because, to satisfy Maxwell's equations, the field $\boldsymbol{B(z)}$ must be an analytic function. Maxwell pointed out the close relationship between the theory of functions of a complex variable and the Laplace equation in two dimensions in chapter XII of his famous work [1], and called methods of the theory of functions of a complex variable, applied for solving 2D problems of electro- and magnetostatics, *much more powerful than any known method applicable in three dimensions*.

Strictly speaking, the representation of the complex *field* rather than the *potential*, in the form $B_y + i B_x$, rather than $B_x + i B_y$, which makes $\boldsymbol{B(z)}$ an analytic function, is not explicitly present in Maxwell's fundamental work; apparently, it appeared much later in Richard A. Beth's works [2]. References to earlier works can be found there as well.

An analytic function is representable by its Taylor series:

$$\boldsymbol{B(z)} = \sum_{n=0}^{\infty} (\boldsymbol{z} - \boldsymbol{z}_0)^n \frac{1}{n!} \frac{d^n \boldsymbol{B}}{d \boldsymbol{z}^n}\bigg|_{\boldsymbol{z} = \boldsymbol{z}_0} \quad (2)$$

which converges to $\boldsymbol{B(z)}$ in some neighborhood for every $\boldsymbol{z}_0$ in the domain. Thus, (1) is a particular representation of the Taylor series for $\boldsymbol{B(z)}$ with $\boldsymbol{z}_0 = 0$ and

$$\boldsymbol{C_n} = \frac{R_{ref}^{n-1}}{(n-1)!} \frac{d^{n-1} \boldsymbol{B}}{d \boldsymbol{z}^{n-1}}\bigg|_{\boldsymbol{z}=0} \quad (3)$$

Neither the reference radius $R_{ref}$ has a physical meaning nor is it related to the radius of convergence. It can be chosen arbitrarily. Usually, it is chosen to be close to the size of the magnet aperture used by the beam, so that the multipole value $\boldsymbol{C_n}$ gives a measure of the influence of the (*n*-1)th derivative of the field on the beam dynamics.

## II. Multipoles due to Distributed Current and Magnetization Density

It can be shown that the current density distribution *j*(*x*, *y*) generates a field with the multipoles

$$\boldsymbol{C_n} = -\frac{\mu_0 R_{ref}^{n-1}}{2\pi} \iint \frac{j(x,y)}{(x+iy)^n} dx\, dy, \quad (4)$$

and the magnetization density distribution $\boldsymbol{m}(x, y) = m_x + i\, m_y$ generates a field with the multipoles

$$\boldsymbol{C_n} = i n \frac{\mu_0 R_{ref}^{n-1}}{2\pi} \iint \frac{\boldsymbol{m}(x,y)}{(x+iy)^{n+1}} dx\, dy \quad (5)$$

Consider the case of a triangular area with vertices located at $z_0, z_1, z_2$, with a current density linearly dependent on *x* and *y*. It is sufficient to consider the case of a current density equal to $j_0$ at $z_0$ and zero at other vertices, since a general case can be obtained as the sum of three terms, each with current density equal to $j_k$ in *k*-th vertex and zero at the others.

To make the following expressions less cumbersome, we introduce the variables $p_k = 1 / z_k$ and denote the triangle area as *S*. With these notations, multipoles can be written in the form

$$\boldsymbol{C_n} = -j_o S \frac{\mu_0 R_{ref}^{n-1}}{2\pi} G_n(p_0, p_1, p_2) \quad (6)$$

For dipole, quadrupole and sextupole coefficients the corresponding functions $G_n$ can be derived as

$$G_1 = -p_0 p_1 p_2 \frac{\left(p_0\left(\ln\left(\frac{p_0}{p_1}\right)(p_2-p_0)^2 + \ln\left(\frac{p_2}{p_0}\right)(p_0-p_1)^2\right) + (p_0-p_1)(p_1-p_2)(p_2-p_0)\right)}{(p_0-p_1)^2(p_1-p_2)(p_2-p_0)^2} \quad (7)$$

$$G_2 = 2p_0^2 p_1 p_2 \frac{\left(\ln\left(\frac{p_0}{p_1}\right)p_1(p_2-p_0)^2 + \ln\left(\frac{p_2}{p_0}\right)p_2(p_0-p_1)^2 + (p_0-p_1)(p_1-p_2)(p_2-p_0)\right)}{(p_0-p_1)^2(p_1-p_2)(p_2-p_0)^2} \quad (8)$$

$$G_3 = -p_0^2 p_1 p_2 \frac{\left(\ln\left(\frac{p_0}{p_1}\right)p_1^2(p_2-p_0)^2 + \ln\left(\frac{p_2}{p_0}\right)p_2^2(p_0-p_1)^2 + p_0(p_0-p_1)(p_1-p_2)(p_2-p_0)\right)}{(p_0-p_1)^2(p_1-p_2)(p_2-p_0)^2} \quad (9)$$

In (7)-(9), the main branch of the complex logarithm should be used in the calculations.

For higher order multipoles, $n > 3$, logarithms drop out of the expressions and the functions $G_n$ can be derived as

$$G_n = \frac{2p_0^2 p_1 p_2}{(n-1)(n-2)(n-3)} \sum_{\substack{j+k+l=n-4 \\ j,k,l \geq 0}} (j+1) p_0^j p_1^k p_2^l \quad (10)$$

where the sum should be taken over all different combinations of $j$, $k$ and $l$.

In a similar setup for the magnetization density linearly dependent on $x$ and $y$, when the magnetization density is equal to $m_0$ at $z_0$ and zero at other vertices, the multipoles are equal to

$$\boldsymbol{C_n} = in\, \boldsymbol{m_o} S \frac{\mu_0 R_{ref}^{n-1}}{2\pi} G_{n+1}(p_0, p_1, p_2) \quad (11)$$

Since any polygonal area can be divided into triangular subdomains, the calculation of integrals over a triangular region is of practical importance.

Integrals (4), (5) can also be calculated exactly for some other cases, but these results have not yet been implemented and tested with any of the popular FEM packages, thus we do not present them here.

## III. IMPLEMENTATION IN ANSYS

To verify the mathematical results obtained, ANSYS [3] had been selected because: a) the license is available to the author; b) the package has the multiphysics capabilities, which allow to account for the multipole calculation deformation due to forces and thermal contraction; c) ANSYS allows access to the database at a low level; and d) the customization procedure is sufficiently well documented.

The multipole calculation is implemented as an external command, named *HRMCALC*, ANSYS itself is not altered. From the ANSYS nodal solution, the magnetic flux density and field are used to calculate the magnetization at the nodes of the elements for which the $\mu_r$ or $BH$ curve is defined. Also, the nodal current density is read out for elements with current density defined. For both sets of elements, nodal displacements are taken into account while calculating multipoles.

The developed program code was compiled as a DLL with Intel Visual Fortran Composer XE 2011 and 2015, and tested with ANSYS 12.0 and 19.0; both versions yield the same result.

The current code version uses only mathematics for triangular areas; quadrangular elements are decomposed into triangles. If the yoke is meshed with quadrangles, this leads to additional errors, albeit very small.

## IV. CASE STUDY

As a case study we present here the calculation of multipoles due to eddy currents in the vacuum chamber of SIS100 dipole magnet, which has ramp rate of up to $4\, T/s$. Since the eddy currents in the chamber are almost constant during the ramp, unless the chamber is heated above $50\, K$, which should never happen, they are most relevant at the beginning, when the main field is minimal and the magnet yoke is far from saturation. In that case, the contribution of eddy currents can be considered independently of the main field. That is, in order to calculate their contribution to field, eddy currents can be calculated analytically from the ramp rate and parameters of the chamber and considered as the only excitation of the magnet.

As the yoke is far from saturation, reference values of multipoles due to eddy currents in the chamber were obtained semi-analytically, using the images method. Then results of the *HRMCALC* command for different mesh densities were compared to the reference values, see TABLE I.

TABLE I. Results of the multipole calculation by the *HRMCALC* command for various mesh densities compared to target values.

| Target values: multipoles by the method of images, [T] @ $R_{ref}$ = 3 cm | | | | | | | |
|---|---|---|---|---|---|---|---|
| $B_1$ | $B_3$ | $B_5$ | $B_7$ | $B_9$ | $B_{11}$ | $B_{13}$ | |
| -1.955E-04 | 2.867E-05 | -2.014E-06 | 8.054E-08 | 1.127E-08 | -1.011E-09 | -3.777E-11 | |

| Number of elements | | | (Calculated in ANSYS) / (Target values) | | | | | | |
|---|---|---|---|---|---|---|---|---|---|
| Total | With current | With $\mu_r$ >1 | $B_1$ | $B_3$ | $B_5$ | $B_7$ | $B_9$ | $B_{11}$ | $B_{13}$ |
| 155728 | 248 | 57183 | 0.9998 | 0.9994 | 1.0090 | 0.9642 | 0.9438 | 1.0242 | 1.7713 |
| 108759 | 207 | 39650 | 0.9998 | 0.9994 | 1.0090 | 0.9636 | 0.9482 | 1.0440 | 2.2929 |
| 69592 | 168 | 25480 | 0.9998 | 0.9994 | 1.0090 | 0.9636 | 0.9491 | 1.0499 | 2.4147 |
| 39186 | 127 | 14430 | 0.9998 | 0.9994 | 1.0090 | 0.9667 | 0.9340 | 0.9842 | 1.1779 |
| 17568 | 87 | 6500 | 0.9998 | 0.9990 | 1.0090 | 0.9685 | 0.9225 | 1.0331 | -0.9452 |

A screenshot of the ANSYS model used is shown in Fig. 1.

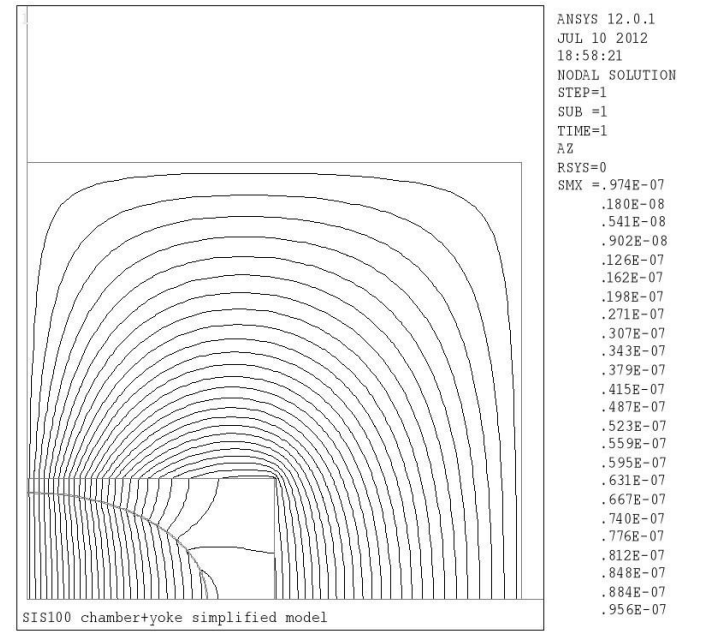

Fig. 1. ANSYS model used in the case study.

## V. Discussion and Conclusion

Commonly, the multipoles of an accelerator magnet in a FEM simulation are calculated by means of a Fourier transform of the field values obtained from the nodal solution on an arc. This requires one to adapt the mesh; the accuracy of the coefficients calculated this way depends strongly on the FEM mesh configuration and simple refinement of the mesh does not always improve accuracy. In some cases, it can be tricky, for example, when the magnet aperture is much smaller, than the dimensions of the coil and yoke.

We derive closed forms for calculating multipoles without using the local FEM solution in the aperture and implemented these results in the ANSYS® add-on. The proposed method of multipoles calculation does not require one to adapt the model for the task, it is suitable for multiphysics simulations, and demonstrates very good stability of results in a wide mesh density range.

The author is ready to provide the ANSYS add-on for accurate calculation of multipoles along with detailed instructions for installing and using the code for those who want to try it.

## Appendix: ~HRMCALC, ANSYS External Command to Calculate Complex Field Expansion Coefficients

### A. Input Parameters

The external ANSYS command *~HRMCALC* calculates the complex field expansion coefficients for 2D magnetic or magneto-structural problems, accounting for the contribution of elements with the current density defined and elements with non-zero magnetization separately. It does not accept any command line arguments, but is controlled by several variables, which may be set in the ADPL script to desired values before issuing the command.

*HARM_OUT* is the number of multipoles to calculate. By default, 15 multipoles, from dipole to 30-pole, will be calculated.
*REF_RAD* is the reference radius. *REF_RAD* = 1[*length*] in the current unit system by default.
*MU_ZERO* is the free space permeability. The MKS system is assumed by default, where $MU\_ZERO = 4\pi \times 10^{-7}\,H/m$.
*X_C_OUT*, *Y_C_OUT* define the origin of the Cartesian coordinate system, parallel to the global the Cartesian coordinate system, where harmonics will be calculated. By default, *X_C_OUT*, *Y_C_OUT* = 0[*length*] in the current unit system, i.e. harmonics will be calculated at the global Cartesian coordinate system origin.
*OFF_STAT* controls the printout of the *~HRMCALC* command. If *OFF_STAT* is not set, or assigned to a value $\leq 0$, then the calculation results along with additional information (number of elements with defined current density and magnetization, status of control variables) will be streamed to the APDL output window (by default) or to a file, if specified by the APDL */OUTPUT* command. If $OFF\_STAT > 0$, the printout is suppressed.

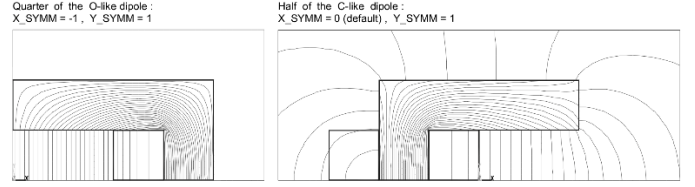

Fig. 2. Examples for setting the symmetry keys *X_SYMM*, *Y_SYMM*

*OFF_DISP*: if this variable is set to a positive value, then *~HRMCALC* calculates harmonics, ignoring displacements (caused by mechanical loads) of elements. This switch is introduced in order to easily assess the importance of the magnet's deformation without needing to solve the model again.
*X_SYMM*, *Y_SYMM* define the symmetry of the problem relatively lines $x = 0$ and $y = 0$ (in the global Cartesian system). By default, it is assumed that the model is complete, i.e. it represents the entire magnet. If $X\_SYMM \neq 0$, then it is assumed that every element has a mirror image with respect to the line $x = 0$. If $X\_SYMM > 0$, then signs of current and magnetization in mirrored elements are chosen so that the magnetic flux is normal to the line $x = 0$, and if $X\_SYMM < 0$, so that the flux is parallel to the line $x = 0$. Similarly, if $Y\_SYMM > 0$, the flux is normal to the line $y = 0$, and if $Y\_SYMM < 0$, the flux is parallel to the line $y = 0$. Certainly, the corresponding boundary conditions must be also applied in the ANSYS model. Examples for *X_SYMM*, *Y_SYMM* settings for one-quarter and one-half inputs of a dipole magnet are shown in Fig. 2.

### B. Output Parameters

*~HRMCALC* places results of the calculation in two APDL arrays: *CURR_HRM* (multipoles, induced by elements with nonzero current density) and *MAGN_HRM* (multipoles, induced by elements with nonzero magnetization). Both arrays have the dimensions [*HARM_OUT*, 2]. The first index in the arrays is the harmonic number (1 - dipole, 2 - quadrupole, etc.), second index has the following meaning: 1 - real ("normal") part of the harmonic, 2 - imaginary ("skew") part. For instance, the element *CURR_HRM*(2,2) gives the value of skew quadrupole for the field induced by currents, and the sum *CURR_HRM*(3,1) + *MAGN_HRM*(3,1) yields the value of normal sextupole for the whole field. To learn more about arrays in APDL, read the ANSYS help on the **DIM* command.

### C. Restrictions and Tips

Present version of *~HRMCALC* cannot treat directly incomplete models with symmetry lines other than $x = 0$ and $y = 0$. In particular, it does not include special switches for multipole symmetries. However, the present version can be used for an accurate calculation of multipoles in simulations of higher order magnets with an incomplete model, if you follow certain recipes when building the model.

- Make the ANSYS input for the smallest non-symmetrical sector of the magnet, that is, the 45° sector for a quadrupole, the 30° sector for a sextupole, 22.5° for an octupole, etc. The sector vertex must be located at the global coor-dinate system origin and one of its boundaries must coincide with the line $x = 0$ or $y = 0$. The model must always include the outermost air area. If it is a coupled, magneto-structural problem, use PLANE13 elements with KEYOPT(1) = 4, or PLANE53 elements, if the problem is purely magnetic.

- set proper conditions on azimuthal boundaries (remember, absence of constraints is equivalent to the flux-normal condition). Set the corresponding value of *X_SYMM* or *Y_SYMM*, depending on which boundary line lies on the coordinate system axis. For instance, if there is a boundary on the line $y = 0$ and the flux is parallel to it, then *Y_SYMM* = -1 should be set. In order to improve the solution precision, recommended not to set any constraint on the flux on the outer radial boundary, but to mesh it with the infinite boundary elements INFIN9.
- if mechanical DOFs are included, constrain UX, UY at the sector vertex and outer radial boundary. On azimuthal boundaries, constrain the displacement, normal to the lines. Set Maxwell surfaces on contours of ferromagnetic areas, which do not coincide with external boundaries.

An example of the quadrupole magnet input, made in accordance with the above recipes, is shown in Fig. 3.

After running the simulation, *~HRMCALC* will calculate harmonics of the field, which are induced by the "doubled" model, completed with the mirror image of actual input relatively to the line for which *X_SYMM* or *Y_SYMM* is set. This "doubled" model represents one sector of a 2*N*-pole magnet, with direct and return currents and correct boundary conditions for calculating the magnetization in the yoke.

In a complete 2*N*-pole magnet, only multipoles $(2k+1)N$, where *k* is a non-negative integer, are allowed by symmetry. The field of one sector of the magnet may contain, in general, all multipoles, the strength of allowed ones is weaker by the factor 2*N*. Therefore, in order to obtain harmonics for the whole 2*N*-pole magnet, allowed elements of the arrays *CURR_HRM* and *MAGN_HRM* must be multiplied by 2*N* (4 for a quadrupole magnet, 6 for a sextupole magnet, etc.) and non-allowed ones should be assigned to zero.

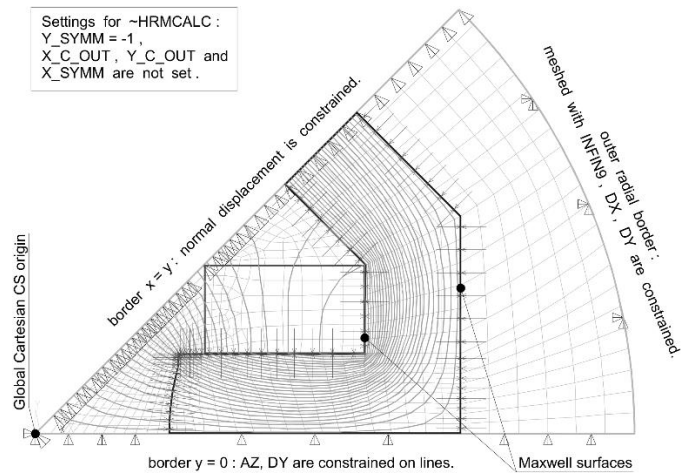

Fig. 3. Quadrupole magnet input example.

### D. General Remarks

*~HRMCALC* works especially well, providing high precision, when the iron yoke is located relatively far from the magnet coil and the field is mostly shaped by conductors, which is typical for high field superconductor magnets. If the field is mostly shaped by the yoke, check results, solving the problem with various mesh configurations.

ACKNOWLEDGMENT

The author would like to thank Markus Kirk for valuable comments and help with preparation of the manuscript.

REFERENCES

[1] James Clerk Maxwell, M.A., "A Treatise on Electricity and Magnetism", Vol.1, Oxford, UK, Clarendon Press, 1873. Available online: https://archive.org/details/treatiseonelectr00maxw, Accessed on: Oct. 28, 2019.
[2] Richard A. Beth, "Evaluation of Current-Produced Two-Dimensional Magnetic Fields", Journal of Applied Physics 40, 4782 (1969). Available online: http://aip.scitation.org/doi/pdf/10.1063/1.1657289?class=pdf, Accessed on: Oct. 28, 2019.
[3] ANSYS® EMag/Mechanical APDL 12.0 and 19.0, ANSYS, Inc.